\documentclass[reprint,prl,twocolumn,superscriptaddress,showpacs,showkeys,floatfix,preprintnumbers]{revtex4-1}

\pdfoutput=1

\usepackage{graphicx}

\usepackage{amssymb}
\usepackage{epsfig}
\usepackage{amssymb}
\usepackage{epsf}
\usepackage{epsfig}
\usepackage{color}
\usepackage{ifpdf}
\usepackage{longtable}
\usepackage{amsmath,amssymb,braket,slashed,mathrsfs}
\usepackage{dcolumn}
\usepackage{hhline}

\newcommand{\nn}{\nonumber}
\newcommand{\ba}{\begin{eqnarray}}
\newcommand{\ea}{\end{eqnarray}}
\newcommand{\be}{\begin{equation}}
\newcommand{\ee}{\end{equation}}
\newcommand{\bd}{\begin{displaymath}}
\newcommand{\ed}{\end{displaymath}}

\newcommand{\MS}{\overline{\mathrm{MS}}}
\newcommand{\lat}{\mathrm{lat}}

\newcommand{\muRI}{\mu_{\mathrm{RI}}}
\newcommand{\RI}{\mathrm{RI}}
\newcommand{\muMS}{\mu_{\overline{\mathrm{MS}}}}

\newcommand{\old}[1]{}

\begin{document}

\title{Exploratory Lattice QCD Study of the Rare Kaon Decay $K^+\to\pi^+\nu\bar{\nu}$}

\date{\today}

\newcommand\cu{Physics Department, Columbia University, New York, NY 10027, USA}
\newcommand\edinb{School of Physics and Astronomy, The University of Edinburgh, Edinburgh EH9 3FD, UK}
\newcommand\soton{Department of Physics and Astronomy, University of Southampton,  Southampton SO17 1BJ, UK}
\newcommand{\pkuphy}{School of Physics, Peking University, Beijing 100871, China}
\newcommand{\innovation}{Collaborative Innovation Center of Quantum Matter, Beijing 100871, China}
\newcommand{\chep}{Center for High Energy Physics, Peking University, Beijing 100871, China}

\author{Ziyuan Bai}\affiliation{\cu}
\author{Norman H. Christ}\affiliation{\cu}
\author{Xu Feng}\affiliation{\pkuphy,\innovation,\chep,\cu}
\author{Andrew Lawson}\affiliation{\soton}
\author{Antonin Portelli}\affiliation{\edinb,\soton}
\author{Christopher T. Sachrajda}\affiliation{\soton}

\collaboration{RBC-UKQCD Collaboration}
\noaffiliation

\begin{abstract}
We report a first, complete lattice QCD calculation of the long-distance
contribution to the $K^+\to\pi^+\nu\bar{\nu}$ decay within the standard model.
This is a second-order weak process involving two four-Fermi operators that is
highly sensitive to new physics and being studied by the NA62 experiment at
CERN.  While much of this decay comes from perturbative, short-distance physics
there is a long-distance part, perhaps as large as the planned experimental
error, which involves nonperturbative phenomena.  The calculation presented
here, with unphysical quark masses, demonstrates that this contribution can be
computed using lattice methods by overcoming three technical difficulties: (i) 
a short-distance divergence that results when the two weak operators approach each other, 
(ii) exponentially growing, unphysical terms that appear in Euclidean,
second-order perturbation theory, and (iii) potentially large finite-volume effects.  A follow-on calculation with physical quark masses and controlled systematic errors will be possible with the next generation of computers.
\end{abstract}



\maketitle

{\em Introduction}. --- An important objective of experimental high-energy physics is the search for
direct and indirect signs of new physics. Complementary to the direct search 
for new particles and forces at high energy, is the search for subtle deviations
from standard model predictions at lower energies. The rare kaon
decays $K\to\pi\nu\bar{\nu}$ are such examples.
As flavor-changing-neutral-current processes, the $K\to\pi\nu\bar{\nu}$ decay 
amplitudes arise as one-loop, electroweak effects.
The small size of one-loop, standard model effects makes 
these decays particularly sensitive to new phenomena. 

These decays are short-distance dominated so that the 
contributions from the strong interactions can be calculated accurately using QCD
perturbation theory. As two of the theoretically cleanest processes, the
$K\to\pi\nu\bar{\nu}$ decays have attracted considerable attention and 
motivate two new experiments.  NA62 at CERN~\cite{fortheNA62:2013jsa}
searches for the $K^+\to\pi^+\nu\bar{\nu}$ decay with a target of
determining the branching ratio to 10\% precision. The KOTO experiment at
J-PARC~\cite{Yamanaka:2012yma}
focuses on the search for the $CP$-violating decay $K_L\to\pi^0\nu\bar{\nu}$ and 
has recently reported the observation of the first candidate event~\cite{Ahn:2016kja}.

Of the two rare kaon decays, the charged decay potentially receives the larger
long-distance contributions.
In fact, the standard model prediction for this decay rate may be enhanced by 6\% when long-distance contributions are included~\cite{Isidori:2005xm}, while
the total uncertainty in the standard model prediction is 10\%~\cite{Buras:2015qea}.  In Ref.~\cite{Christ:2016eae} we have presented
a method using lattice QCD that allows a first-principles calculation of these 
long-distance contributions with controlled errors. Here we apply 
this approach, carrying out a complete, exploratory lattice QCD calculation
of the long-distance contributions to the $K^+\to\pi^+\nu\bar{\nu}$ decay with unphysical quark masses.

The methods used here are closely related to those developed 
earlier to compute other, second-order electroweak effects, specifically
the $K_L$-$K_S$ mass difference~\cite{Christ:2012se, Bai:2014cva}
and the long-distance contributions to the indirect $CP$-violating parameter $\epsilon_K$~\cite{Bai:2015xxx, Bai:2016gzv}.  This work is also part of a 
larger effort that includes the lattice QCD calculation of the rare kaon 
decays $K\to\pi\ell\bar{\ell}$~\cite{Christ:2015aha,Christ:2016mmq}.

{\em Formulation}. --- As explained in Ref.~\cite{Christ:2016eae}, the  $K^+\to\pi^+\nu\bar{\nu}$ decay amplitude is conventionally expressed as the sum of top- and charm-quark contributions. The long-distance part of interest appears in the charm quark contribution which can be written as the matrix element of a combination of bilocal and local operators of the form
\begin{eqnarray}
\hskip -0.2 in \mathcal{O}(y) = \sum_{A,B}\int d^4x\,T[C_A Q_A(x)\,C_B Q_B(y)] + C_0 Q_0(y),
\label{eq:H_eff_2nd}
\end{eqnarray}
where $T$ indicates a time-ordered product and the local operator
$Q_0=\sum_{\ell=e,\mu,\tau}(\bar{s}d)_{V-A}(\bar{\nu}_\ell\nu_\ell)_{V-A}$.  The
Wilson coefficients $C_S(\mu)$ contain short-distance information from the $W$
scale down to the lower energy scale $\mu$ at which the operators $Q_S$ are
renormalized.  The $Q_S$ with $S=A,B$ are seven, four-Fermi operators which
enter the first-order, weak Hamiltonian density, $\mathcal{H}_{\mathrm{eff}} =
\sum_S \widetilde{C}_S Q_S$, where $\widetilde{C}_S$ is the product of $C_S(\mu)$, a CKM matrix element and other conventional factors.  

When two first-order operators are multiplied in such a second-order
calculation, {\it e.g.}, $Q_A(x) Q_B(0)$, new singularities appear as $x\to0$. The counterterm $C_0Q_0(0)$ removes these singularities and reproduces the physical amplitude.   For sufficiently large $\mu$, {\it e.g.} $\mu = 3 $ GeV, the coefficient $C_0$ can be determined in perturbation theory.  

If the bilocal operator in Eq.~\eqref{eq:H_eff_2nd} is renormalized at such a
large scale $\mu$, then the short-distance physics from the scale of the $W$
mass down to $\mu$ will be represented by the local operator $C_0Q_0$ which,
enhanced by $\ln(M_W^2/\mu^2)$, is expected to give the largest contribution and
is readily evaluated because $C_0$ is known from perturbation theory and
$\langle\pi|Q_0|K\rangle$ can be determined from the measured $K_{\ell3}$ form
factor $F_+$.  It is the bilocal operator in Eq.~\eqref{eq:H_eff_2nd} which is
the focus of this Letter.

In the conventional treatment~\cite{Buras:2015qea} the bilocal operator is also approximated by $Q_0$ multiplied by a perturbative Wilson coefficient $r_{AB}(\mu)$, obtained by integrating out the charm quark.  Combining $r_{AB}(\mu)$ with $C_0(\mu)$, one determines the total Wilson coefficient for $Q_0$, written as
\ba
P_c^{\mathrm{PT}}=\frac{1}{\lambda^4}\frac{\pi^2}{M_W^2}\Bigl(\sum_{A,B}C_A(\mu)C_B(\mu)r_{AB}^{\MS}(\mu)+C_0(\mu)\Bigr),\ \ \
\label{eq:PT}
\ea
where $\lambda$ is the CKM matrix element $|V_{us}|$ and the label PT has been introduced to identify a perturbative result. $P_c^{\mathrm{PT}}$ has been calculated in NNLO QCD perturbation theory, giving $P_c^{\mathrm{PT}}=0.365(12)$ where the error reflects the dependence on $\mu$~\cite{Buras:2015qea}.  A correction to $P_c^{\mathrm{PT}}$, which estimates up-quark and other long-distance effects suppressed by $(\Lambda_{\mathrm{QCD}}/m_c)^2$, is written as $\delta P_{c,u}=0.04(2)$~\cite{Isidori:2005xm}.

The errors in this conventional treatment of the bilocal operator are expected to be a few percent but are difficult to estimate or to reduce.  Here we use lattice QCD to provide a first-principles and systematically improvable calculation of the contribution of this bilocal operator.

In the standard perturbative calculation which determines $\mathcal{O}(y)$, the
Wilson coefficients $C_S(\mu)$ and $C_0(\mu)$ are computed and the local
operators and bilocal operator products are renormalized in the 
modified minimal subtraction ($\MS$) scheme.  As described in greater detail in Ref.~\cite{Christ:2016eae}, we relate these $\MS$ operators to lattice operators by using an intermediate, regularization-independent
symmetric momentum (RI/SMOM) scheme~\cite{Sturm:2009kb,Christ:2016eae}, illustrated for $Q_A$ and $Q_B$ by the equation
\begin{eqnarray}
    \left\{\int
    d^4x\,T\left[Q_A^{\MS}(x)\,Q_B^{\MS}(0)\right]\right\}^{\MS}_\mu&& 
\label{eq:bilocal_renorm}\\
&& \hskip -1.7 in =Z_A^{\lat\to\MS}Z_B^{\lat\to\MS}\left\{\int
    d^4x\,T\left[Q_A^{\lat}(x)Q_B^{\lat}(0)\right]\right\}^{\lat}_a
    \nonumber\\
    &&\hskip -1.5 in -Z_A^{\RI\to\MS}Z_B^{\RI\to\MS}X_{AB}^{\lat\to\RI}(\muRI,a)\left\{Q_0(0)\right\}^{\RI}_{\muRI}
\nonumber \\
&& \hskip -1.5 in+Y_{AB}^{\RI\to\MS}(\mu,\muRI)\left\{Q_0(0)\right\}^{\MS}_{\mu},
\nonumber
\end{eqnarray}
where the renormalization factors $Z_{A(B)}^{\mathcal{S}\to\mathcal{S}'}$
convert $Q_{A(B)}$ from scheme $\mathcal{S}$ to scheme  $\mathcal{S}'$, assuming they are multiplicatively renormalized, and $a$ is the lattice spacing.  To handle the singularity at $x=0$ in the product $Q_A^{\lat}(x)Q_B^{\lat}(0)$, we introduce the $Q_0$ term. By adding the counterterm $X_{AB}^{\lat\to\RI}(\muRI,a)\{Q_0\}^{\RI}_{\mu_\RI}$, we first convert the simple bilocal product of individually renormalized $\RI$ operators into a bilocal operator renormalized in the RI/SMOM scheme. The Wilson coefficient $X_{AB}^{\lat\to\RI}(\muRI,a)$ can be determined nonperturbatively by imposing the RI/SMOM renormalization condition, described below, at a scale $\muRI$.  In the second step, we use QCD perturbation theory to determine the $Y_{AB}^{\RI\to\MS}(\mu,  \muRI)\{Q_0\}^{\MS}_\mu$ term, which converts the RI/SMOM bilocal operator to a $\MS$ operator, renormalized at the scale $\mu$.  The use of perturbation theory requires $\mu,\;\muRI\gg \Lambda_{\mathrm{QCD}}$. Greater detail is given in Ref.~\cite{Christ:2016eae}.

{\em Lattice ensemble}. --- We use the $16^3\times32$, $2+1$ flavor, domain wall
fermion ensemble, with $a^{-1}=1.729(28)$ GeV and a fifth-dimensional extent of
$L_s=16$ generated by the RBC and UKQCD Collaborations~\cite{Blum:2011pu}.  This ensemble has a residual mass  $m_{\mathrm{res}}a=0.00308(4)$ and pion and kaon masses of $M_\pi=421(1)(7)$ MeV and $M_K=563(1)(9)$ MeV.  We use a valence charm mass, $m_c a=0.330$, giving an $\MS$ mass $m_c^{\MS}(\mbox{2 GeV})=863(24)$ MeV.  We analyze 800 gauge configurations, each separated by 10 molecular dynamics time units.

We work in the kaon rest system and describe the $\pi^+\nu\bar{\nu}$ final state using the Dalitz variables $s=-(p_K-p_\pi)^2$ and $\Delta=(p_K-p_\nu)^2-(p_K-p_{\bar{\nu}})^2$.  Since $M_\pi\approx420$ MeV, the allowed, final-state momenta lie in a narrow region. Assuming little variation across this region,
we use the single momentum choice $(\Delta,s)=(0,0)$ by fixing the pion spatial momentum $\vec{p}_\pi=(0.0414,0.0414,0.0414)/a$ so that the neutrino and antineutrino move in the opposite direction, each carrying the momentum $-\vec{p}_\pi/2$.  The pion's spatial momentum is fixed by imposing twisted boundary conditions on the down valence quark.

   \begin{figure}
   \centering
        \shortstack{
	\shortstack{\includegraphics[width=.2\textwidth]{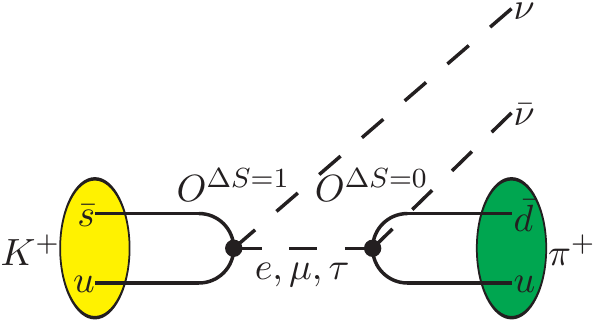}\\ (a) Type 1}
	\hspace{0.3cm}
        \shortstack{\includegraphics[width=.2\textwidth]{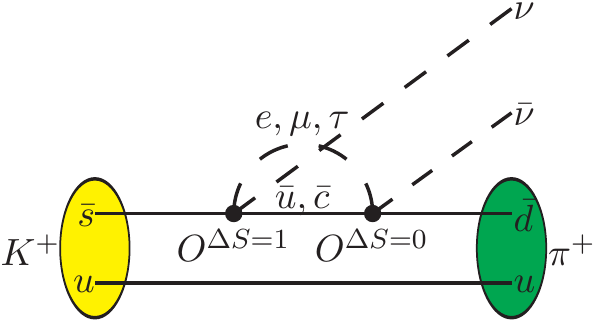}\\ (b) Type 2}
	\\$W$-$W$ diagram}\\
	\ \\
	\shortstack{
    \shortstack{\includegraphics[width=.2\textwidth]{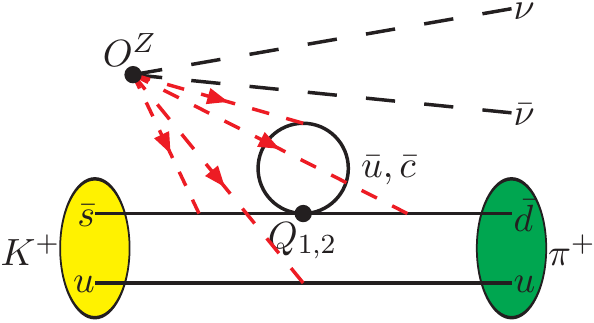}\\ (c) With closed loop}
    \hspace{0.3cm}
    \shortstack{\includegraphics[width=.2\textwidth]{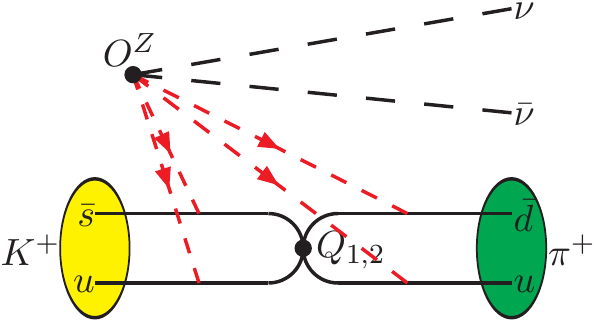}\\ (d) Without closed loop}
    \\Connected $Z$-exchange diagram}\\
    \ \\
    \shortstack{
    \shortstack{\includegraphics[width=.2\textwidth]{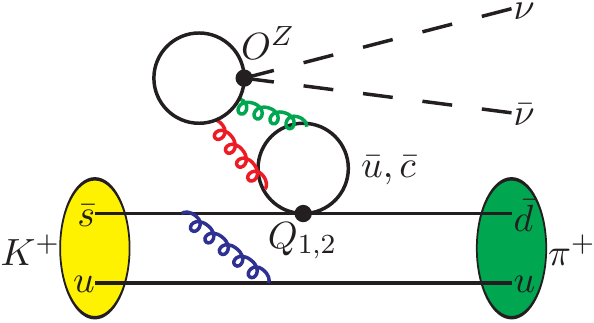}\\ (e) With closed loop}
    \hspace{0.3cm}
    \shortstack{\includegraphics[width=.2\textwidth]{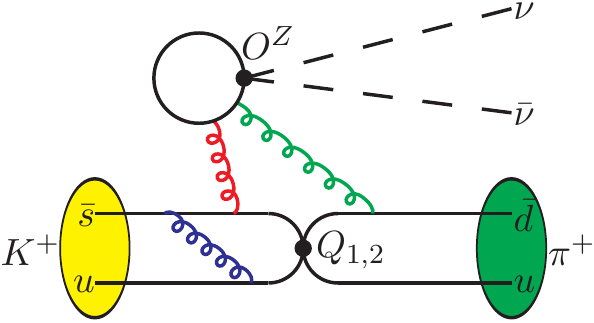}\\ (f) Without closed -loop}
    \\ Disconnected $Z$-exchange diagram}
   \caption{From top to bottom: quark and lepton contractions for $W$-$W$, connected and disconnected $Z$-exchange diagrams.  The four dotted arrows point to possible locations for the $Z$-exchange vertex.  The operator labels are defined in Ref.~\cite{Christ:2016eae}.  A few, illustrative gluon lines are also shown.}
   \label{fig:contraction}
   \end{figure}

{\em Bilocal operator}. --- The Feynman diagrams corresponding to the matrix
element of the bilocal operator in Eq.~\eqref{eq:H_eff_2nd} are shown in
Fig.~\ref{fig:contraction}.   We use Coulomb-gauge-fixed wall sources for the
valence quarks propagators joined to the initial kaon and final pion states. For
the diagrams (a), (b), and (d), which do not contain a closed quark loop, we
treat the two weak interaction vertices asymmetrically.  One is evaluated at a
fixed point, which is used as the source for the internal quark lines connected to that operator.  The second operator acts as the sink for all the propagators joined to it and is summed over the desired space-time subvolume.  For higher precision we average over time translations, calculating these wall- and the point-source propagators for all 32 time slices.  In the $W$-$W$ diagrams in Fig.~\ref{fig:contraction}, we also exchange the source and sink locations between the two weak operators and average over both choices.

Each internal lepton propagator is that of an overlap fermion with an infinite
time extent and a physical lepton mass.  For the $Z$-exchange diagram where the
decay involves a four-quark operator and a two-quark-two-neutrino operator, both
operators can generate a closed quark loop.  Thus, we need to calculate the
diagonal element of the light, strange, and charm quark propagators $D^{-1}(x,x)$ for all space-time positions $x$. This is done by using 32 random, space-time volume sources for each quark flavor.  We perform a complete calculation, including all connected and disconnected diagrams.

In a lattice QCD calculation, the matrix element of the time-integrated bilocal operator appearing in Eq.~\eqref{eq:H_eff_2nd} is evaluated in Euclidean space.  As in the case of the calculation of the $K_L$-$K_S$ mass difference~\cite{Christ:2012se}, this matrix element can be related to the second-order amplitude of interest if a sum over intermediate states is inserted and the integration over Euclidean time performed:
\begin{eqnarray}
\int_{-T_a}^{T_b} d x_0 \langle \pi^+\nu\overline{\nu}| T\left\{H_A(x_0)H_B(0)\right\}|K^+\rangle && 
\label{eq:int_sum} \\
&&\hskip -2.4 in
=  \sum_n \Biggl\{ \frac{\langle \pi^+\nu\overline{\nu}|H_A|n\rangle\langle n|H_B|K^+\rangle}{E_n-E_K}
                           \left(1-e^{(E_K-E_n)T_b}\right) 
\nonumber \\
&&\hskip -2.1 in + \frac{\langle \pi^+\nu\overline{\nu}|H_B|n\rangle\langle n|H_A|K^+\rangle}{E_n-E_K}
                           \left(1-e^{(E_K-E_n)T_a}\right)
\nonumber \Biggr\},
\end{eqnarray}
where we have replaced the local operators in Eq.~\eqref{eq:H_eff_2nd} by those
integrated over space: $H_S(x_0) = \int d^3 x Q_S(\vec x, x_0)$.  The unphysical
$e^{(E_K-E_n)T_{a(b)}}$ terms in the second and third lines of this equation
vanish for large $T_{a(b)}$ for intermediate states more energetic than the
kaon.   However, these terms grow exponentially with increasing integration
range if $E_n < E_K$.  These are calculated separately and their contributions
removed; see Refs.~\cite{Christ:2012se, Christ:2016eae, Bai:2015xxx, Christ:2015aha}.

A second difficulty implied by Eq.~\eqref{eq:int_sum} is the possibility of a large contribution caused by a vanishing denominator when a finite-volume intermediate-state energy $E_n$ approaches $E_K$~\cite{Christ:2016eae}.  Such behavior is a well-understood finite-volume effect and a complete correction can be applied~\cite{Christ:2015pwa}.  Thus, we must pay special attention to three states $|n\rangle=|\ell^+\nu\rangle$, $|\pi^0\ell^+\nu\rangle$ and $|(\pi^+\pi^0)^{I=2}\rangle$ and calculate all the transition amplitudes for $K^+\to |n\rangle$ and
$|n\rangle\to|\pi^+\nu\bar{\nu}\rangle$ both to remove the exponentially growing
terms and to estimate finite-volume effects. 

Because of the $V-A$ structure of the weak interactions and the vanishing mass of the final-state neutrinos, the bilocal matrix element can be written as the product of a scalar amplitude and the spinor quantity $\bar{u}(p_\nu){\slashed p}_K(1-\gamma_5) v(p_{\bar{\nu}})$, as is shown in Ref.~\cite{Christ:2016eae}. For the $W$-$W$ diagrams this scalar amplitude is written as $F_{WW}(\Delta,s)$. 

For the $Z$-exchange diagrams the scalar amplitude is given by a $K_{\ell3}$-like form factor $F_+^Z(s)$.   For massless neutrinos, a second form factor, $F_-^Z(s)$ does not
contribute.  We compute $F_+^Z(0)$ for the connected diagrams as described earlier and $F_0^Z(s)=F_+^Z+s F_-^Z/(M_K^2-M_\pi^2)$ for both the connected and disconnected parts at $\vec p_K=\vec p_\pi=0$ and $s=s_{\mathrm{max}}=(M_K-M_\pi)^2$.  We calculate $F_0^{Z,\mathrm{disc}}(s_{\mathrm{max}})$ instead of $F_+^{Z,\mathrm{disc}}(0)$ to avoid using twisted momenta for the disconnected graphs and expect this to have a small effect since $s_{\mathrm{max}}/(M_K^2-M_\pi^2)=0.14 \ll 1$ and $F_+^{Z,\mathrm{conn}}(0) \approx F_0^{Z,\mathrm{conn}}(s_{\mathrm{max}})$ as seen in Table~\ref{table:matrix_element}.

\begin{table}[ht]
\centering
\begin{tabular}{lD{.}{.}{8}cD{.}{.}{12}c}
\hhline{==~==}
\multicolumn{2}{c}{scalar amplitude} && \multicolumn{2}{c}{contribution from state $|n\rangle$} \\ 
\hhline{--~--}
$F_{WW}$ type 1 & -1.118(26) && -1.138(4) & $|\ell^+\nu\rangle$        \\
$F_{WW}$ type 2 & 9.29(14)     &&0.657(5) & $|\pi^0\ell^+\nu\rangle$\\
$F^{Z,\mathrm{conn}}_+(0)$ & 2.133(32)  &&  \multicolumn{2}{c}{$\cdots$}  \\
$F^{Z,\mathrm{conn}}_0(s_{\mathrm{max}})$ & 2.109(25) && 0.1526(10)  & $|(\pi^+\pi^0)^{I=2}\rangle$ \\
$F^{Z,\mathrm{disc}}_0(s_{\mathrm{max}})$ & 0.060(12) &&
\multicolumn{2}{c}{$\cdots$}  \\
\hhline{--~--}
\end{tabular}
\caption{Resulting scalar amplitudes for the $W$-$W$ and $Z$-exchange diagrams. All the results are shown in lattice units (in units of $10^{-2}$). The scalar amplitude $F_{WW}$ is evaluated at $(\Delta,s)=(0,0)$, $F_+^Z$ at $s=0$ and $F_0^Z$ at $s=(M_K-M_\pi)^2$.}
\label{table:matrix_element}
\end{table}

Our results for the various components of the scalar amplitude are shown in
Table~\ref{table:matrix_element}. For the $W$-$W$, type 1 diagram, the dominant
contribution to $F_{WW}$ comes from the lowest intermediate state
$|\ell^+\nu\rangle$. The type 2 diagram yields a much larger contribution than type 1.
Since it involves a fermion loop, the dominant contribution comes from
short distances where new divergences appear and a short-distance correction is
required. The $|\pi^0\ell^+\nu\rangle$ intermediate state contributes only about
8\% to $F_{WW}$.  

For the $Z$-exchange diagram, the $|(\pi^+\pi^0)^{I=2}\rangle$ state contributes
about 7\%. Although with $M_\pi\approx420$ MeV, the contribution of this state
to an exponentially growing, unphysical term or to finite-volume corrections is irrelevant, this state could cause significant systematic effects for a calculation at the physical pion mass.

As described above, we have also evaluated the disconnected diagrams.  Although
the result is noisy, the size of the disconnected diagrams is only 3\% of the
connected diagrams. Thus, including the disconnected diagrams will not affect the
statistical precision of our result.

{\em Local operator}. --- The matrix element of the local operator $Q_0^{\mathrm{lat}}$ is related to the matrix element of the conserved vector current between a kaon and pion and can be determined from $K_{\ell3}$ decay without reference to lattice QCD.  (Of course, for our unphysical kinematics a lattice calculation is needed.)  Here we will focus on the coefficient of this operator, specifically the contributions to this coefficient from the terms in the third and fourth lines of Eq.~\eqref{eq:bilocal_renorm}: the terms that renormalize the bilocal lattice operator discussed above.

As discussed in detail in Ref.~\cite{Christ:2016eae}, the coefficient
$X_{AB}^{\lat\to\RI}(\muRI,a)$, which converts the lattice bilocal operator into one defined in the RI/SMOM scheme can be determined from a nonperturbative
calculation of an off-shell, Landau-gauge-fixed Green's function of five operators: the four quark fields $\bar{s}$, $d$, $\nu$ and $\bar{\nu}$ carrying nonexceptional, external Euclidean momenta and the sum of the operators appearing in the second and third lines of Eq.~\eqref{eq:bilocal_renorm}.  We use the external four-momenta:
\ba
\label{eq:extern_mom}
&&p_{\bar{s}}=(\xi,\xi,0,0),\quad p_d=(\xi,0,\xi,0),
\nn\\
&&p_{\bar{\nu}}=(0,-\xi,0,-\xi),\quad
p_\nu=(0,0,-\xi,-\xi),
\label{eq:RISMOM}
\ea
where $-p_{\bar{s}}$, $p_d$, $-p_{\bar{\nu}}$ and $p_\nu$ are incoming.
The RI/SMOM scale is $\muRI^2=p_f^2=2\xi^2$, for $f=\bar{s}$, $d$, $\nu$, $\bar{\nu}$.  The spin and color indexes of the external fermion lines are contracted in the same fashion as those in the operator $Q_0$.   The coefficient $X_{AB}^{\lat\to\RI}(\muRI,a)$ is determined by requiring that the Green's function described above vanishes for the momenta in  Eq.~\eqref{eq:RISMOM} and a specific choice of $\muRI$.  The resulting RI/SMOM-renormalized, bilocal operator now has a well-defined continuum limit.  In this way we obtain $X_{AB}^{\lat\to\RI}(\muRI, a)$ for 1 GeV $\le\muRI\le$ 4 GeV.  

Next we calculate the coefficient $Y_{AB}^{\RI\to\MS}(\mu,\muRI)$ needed to
convert the RI-renormalized operator to $\MS$ renormalization. This can be done
directly from Eq.~\eqref{eq:bilocal_renorm} by evaluating both sides at the
external momenta specified in Eq.~\eqref{eq:RISMOM} at the scale $\muRI$.  The
left-hand side is evaluated in perturbation theory. On the right-hand side the
first and second lines are, in principle, nonperturbative but cancel exactly because of the definition of the RI/SMOM scheme.  The remaining term, $Y_{AB}^{\RI\to\MS}(\mu,\muRI)$, is thus determined.  For simplicity, we choose  $\mu=\muRI$ and evaluate $Y$ perturbatively at one-loop.  Knowing the Wilson coefficients $X$ and $Y$, the contribution of the local operator $Q_0$ is easily computed.

\begin{figure}
\centering
\includegraphics[width=.48\textwidth]{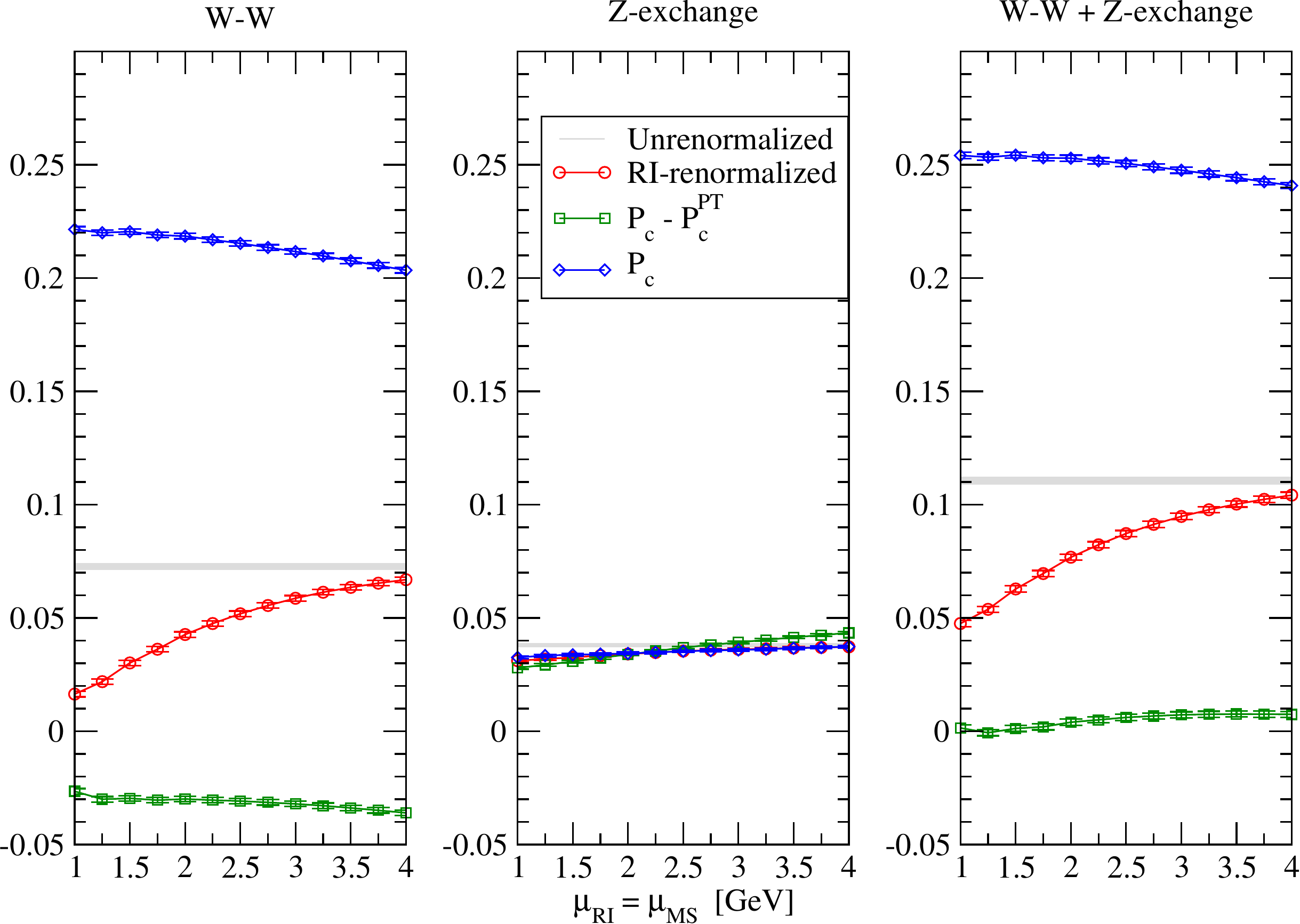}
\caption{$W$-$W$ and $Z$-exchange diagram results, and their total, shown from left to the right.  The gray bands show the amplitude, normalized as in Eq.~\eqref{eq:PT}, from the unrenormalized, bilocal operator. The red circles indicate the RI-renormalized, bilocal contribution. The blue diamonds give the total charm contribution $P_c$ while the green squares show the difference between the lattice and perturbative results, $P_c - P^{\mathrm{PT}}_c$. }
\label{fig:Pc}
\end{figure}

{\em Results}. --- Since we use an unphysical value for the charm quark mass,
$m_c^{\MS}(\mbox{2 GeV})=863(24)$ MeV, we reevaluate $P_c^{\mathrm{PT}}$ of
Eq.~\eqref{eq:PT} using this unphysical value and the NNLO formulas of Ref.~\cite{Buras:2006gb}.  Our results, including statistical errors, are shown in Fig.~\ref{fig:Pc}.   Here $P_c$ gives the complete charm contribution to the $K^+\to\pi^+\nu\bar\nu$ decay, normalized so that the decay amplitude is the matrix element of the operator 
$\alpha G_F \lambda^5/(2\pi\sqrt{2}\sin^2\theta_W) P_c Q_0$ where $\alpha$ is the fine structure constant, $G_F$ the Fermi constant and $\theta_W$ the Weinberg angle.  This description neglects the dependence of the decay amplitude on the Dalitz variables $s$ and $\Delta$, which will be small for our kinematics. 

We show results from the $W$-$W$ diagrams, the $Z$-exchange diagrams and their total in the left, center and right panels.  First, as the gray band, we plot the lattice matrix element of the bilocal operator with only the multiplicative renormalization of the individual four-Fermi operators included.  Second, as red circles, we show the matrix element of the bilocal operator, now normalized in the RI/SMOM scheme.  The short distance subtraction has introduced a dependence on $\muRI=\muMS$.  (The local operators are renormalized at the fixed scale $\muMS=2.15$ GeV.)  Next, we plot our complete result, $P_c$ as blue diamonds.  Finally, as green squares we show $P_c - P_c^{\mathrm{PT}}$,  the difference  between our complete lattice result and the result of perturbation theory $P_c^{\mathrm{PT}}$ described above.

The results from our exploratory lattice calculation with unphysical charm, down and up quark masses are:
   \ba
   P_c&=&0.2529(\pm13)(\pm32)(-45) \nn\\
   P_c - P^{\mathrm{PT}}_c &=&0.0040(\pm13)(\pm32)(-45),
\label{eq:results}
   \ea
where the first uncertainty is statistical and the second is the systematic
uncertainty arising from the scale dependence as $\mu$ varies between 1 and 3
GeV.  The third quantity is an estimate of finite-volume errors.  With
$M_\pi\approx420$ MeV, only the $|\pi^0e^+\nu\rangle$ state can cause such
effects, whose size is determined from the formulas in
Ref.~\cite{Christ:2016eae}.  The use of 800 configurations and unphysically
heavy up and down quarks yields a subpercent statistical error for $P_c$.  The
small size of $P_c - P^{\mathrm{PT}}_c$ results from a large cancellation
between the $W$-$W$ and $Z$-exchange amplitudes. It is important to determine
whether such a large cancellation persists for physical quark masses since, for
example, if only the $W$-$W$ piece were present the predicted branching ratio would decrease by 6\%.  

{\em Conclusion} --- The rare decay $K^+\to\pi^+\nu\bar{\nu}$ is a promising process to reveal new physics both because of its small size and the accuracy with which the dominant, short-distance parts can be computed in the standard model.  While the top quark alone contributes 50\% of the branching ratio, amplitudes containing the much lighter charm quark do appear in the other 50\%.  However, at leading order most of this charm contribution comes from the short-distance-dominated logarithm, $\ln(M_W^2/m_c^2) \approx 8.4$, suggesting that long-distance effects may give only 10\% of the charm contribution or 5\% of the branching ratio.

Since such estimates are necessarily uncertain [for example, the
    $\ln(M_W^2/m_c^2)$ piece is reduced by a factor of 2 when all leading
logarithms are included] and the NA62 experiment plans to measure this branching ratio to 10\%, a direct lattice QCD calculation of these long-distance effects is well motivated.  The exploratory calculation presented here demonstrates that this is possible.

Because of our unphysical quark masses, it is premature to compare the
difference between our result and the perturbative
calculation~\cite{Buras:2015qea} given in Eq.~(\ref{eq:results}) with the  phenomenological, long-distance correction $\delta P_{c,u}=0.04(2)$ of Ref.~\cite{{Isidori:2005xm}}.  However, the techniques presented here can be directly applied to a future, realistic calculation. We expect that within four years, when adequate resources become available, an accurate lattice calculation with controlled systematic errors will be possible.

{\em Acknowledgments}. --- We thank our colleagues in the RBC and UKQCD
Collaborations for many helpful discussions.  Z.B., N.C., and X.F. were
supported in part by U.S. DOE Grant No. De-SC0011941, while A.P. and C.T.S. were
supported in part by UK STFC Grant No. ST/L000296/1.   A.P. also received
support from UK STFC Grant No. ST/L000458/1.   A.L. is supported by an EPSRC Doctoral Training Centre Grant No. EP/G03690X/1.

\bibliography{K2pivv}
\bibliographystyle{h-physrev}

\end{document}